\documentclass[a4paper,12pt]{article}

\usepackage{graphicx}
\usepackage{citesort}
\usepackage{epsfig}

\makeatletter

\@addtoreset{equation}{section}
\makeatother
\def\lsim{\;\raise0.3ex\hbox{$<$\kern-0.75em\raise-1.1ex\hbox{$\sim$}}\;}
\def\gsim{\;\raise0.3ex\hbox{$>$\kern-0.75em\raise-1.1ex\hbox{$\sim$}}\;}

\def\ben{\begin{enumerate}}  \def\een{\end{enumerate}}
\def\bit{\begin{itemize}}    \def\eit{\end{itemize}}
\def\beq{\begin{equation}}   \def\eeq{\end{equation}}
\def\ba{\begin{array}}       \def\ea{\end{array}}
\def\bea{\begin{eqnarray}}   \def\eea{\end{eqnarray}}
\def\nn{\nonumber}

\def\LL1{\mathrm{L}_1}
\def\L2{\mathrm{L}_2}

\begin{document}
\baselineskip 18pt

\begin{titlepage}
\renewcommand{\thefootnote}{\fnsymbol{footnote}}
\setcounter{footnote}{0}

\begin{flushright}
LPT Orsay 07-88 \\
\end{flushright}

\begin{center}
\vspace{3cm}
{\Large\bf Updated Constraints from B Physics on the MSSM and the
NMSSM} \\
\vspace{2cm}

{\bf Florian Domingo\footnote{email: domingo@th.u-psud.fr} and 
Ulrich Ellwanger\footnote{email: ellwanger@th.u-psud.fr}} \\
Laboratoire de Physique Th\'eorique\footnote{Unit\'e mixte de Recherche
-- CNRS -- UMR 8627} \\
Universit\'e de Paris XI, F-91405 Orsay Cedex, France
\vspace{2cm}
\end{center}

\begin{abstract}
We update constraints from B physics observables on the parameters of
the MSSM and the NMSSM, combining them with LEP constraints. Presently
available SM and Susy radiative corrections are included in the
calculations, which will be made public in the form of a Fortran code.
Results for the $\tan\beta$ and $M_{H^\pm}$ dependence of\break 
$BR(\bar{B} \to X_s \gamma)$ are presented, as well as constraints on
the NMSSM specific case of a light CP odd Higgs scalar. We find that
the latter are essentially due to $BR(\bar{B}_s\to \mu^+ \mu^-)$, but
they do not exclude this possibility.
\end{abstract}

\end{titlepage}

\renewcommand{\thefootnote}{\arabic{footnote}}
\setcounter{footnote}{0}

\section{Introduction} 

It is well known that rare decays and/or oscillations of $B$-Mesons
impose constraints on the para\-meter space of models Beyond the
Standard Model (BSM): BSM contributions are not necessarily suppressed,
once the dominant contributions both in the SM and BSM arise from
loop diagrams (or are even absent in the SM).

Recently, considerable progress has been made both on the experimental
side (such as improved measurements of small branching ratios) and on
the theoretical side, i.e. improved evaluations of SM predictions
and BSM contributions.

The purpose of the present paper is to study the resulting constraints
on the parameter space of supersymmetric extensions of the standard
model, both in the MSSM and the NMSSM, from $BR(\bar{B} \to X_s
\gamma)$, $\Delta M_s$, $\Delta M_d$, $BR(\bar{B}_s \to \mu^+ \mu^-)$
and $BR(\bar{B}^+ \to \tau^+ \nu_\tau)$. In the MSSM, similar analyses
have recently been performed in \cite{carena,carena2,IsP,ellis}
(see also refs. \cite{blanke1,ball,blanke2,freitas,isidori,alt} for
recent discussions within the Minimal Flavour Violating MSSM).

In \cite{carena,carena2} the new experimental B physics results
have been used to constrain the parameter space of the MSSM. In
\cite{IsP} it has been argued, that the new results on $BR(\bar{B}^+
\to \tau^+ \nu_\tau)$ are evidence for BSM contributions. A general
$\chi^2$ fit has been performed in \cite{ellis} in the context of the
CMSSM (with universal Susy-breaking terms at the GUT scale) and the
NUHM (with non-universal Higgs mass terms), together with constraints
on the dark matter relic density.

One purpose of the present paper is to consider constraints from
$BR(\bar{B} \to X_s \gamma)$ on the NMSSM. Our result
is that the NMSSM specific effects on $BR(\bar{B} \to X_s \gamma)$ are
rather weak: in the NMSSM the charged Higgs mass squared receives (at
tree level) a negative contribution relative to the MSSM which lowers
its mass somewhat; once the result of $BR(\bar{B} \to X_s \gamma)$
is plotted against $M_{H^\pm}$, no difference between the MSSM and the
NMSSM remains visible, however. Two loop corrections (relevant at large
$\tan\beta$) are sensitive to the neutralino sector which includes the
singlino in the NMSSM; we find, however, that even for relatively large
singlino -- MSSM-like-neutralino mixings the NMSSM specific numerical
effect on $BR(\bar{B} \to X_s \gamma)$ is numerically negligible.
(Combined constraints on the parameter space of the NMSSM from LEP, the
dark matter relic density and B physics -- but without the recent
developments in B physics -- have been investigated previously in
\cite{cerdeno,belanger}.)

Note that in the general MSSM, LEP constraints on the lightest Higgs
mass impose $\tan\beta \gsim 3$ (or $\tan\beta \gsim 10$ in the CMSSM).
In the NMSSM (and the CNMSSM), LEP constraints on Higgs masses and
couplings allow for rather low values of $\tan\beta$ \cite{eh1,eh2};
here $\tan\beta$ can be as low as 1.5. Our results for $BR(\bar{B} \to
X_s \gamma)$ for low values of $\tan\beta$ (which have not been
considered in \cite{carena,carena2,IsP,ellis}) are thus specific to the
NMSSM, although the results in the MSSM (without LEP constraints) would
have been the same.

In the NMSSM, important new contributions to B physics observables can
originate from the presence of a relatively light CP odd Higgs boson
\cite{lighta1,lighta2,hiller,lighta3,lighta4,lighta5,lighta6,lighta7},
which can also be consistent with the dark matter relic density
\cite{cerdeno,belanger}, and which can contribute significantly
via\break 
s-channel single and double penguin diagrams to $B$ physics processes
even for small $\tan\beta$. Constraints from B physics  observables on
this region of the parameter space of the NMSSM will be discussed in
section~5.

Our numerical results are obtained with the help of a Fortran code,
that will be made public as a part of the NMSSMTools package
\cite{nmtools}. It allows us to combine the constraints on the
parameter space from $B$ physics with constraints on the Higgs sector
from LEP. (In the MSSM, subroutines that compute $B$ physics
observables are included in FeynHiggs \cite{feynhiggs}, Suspect
\cite{suspect}, MicrOmegas \cite{micro1,micro2} and Spheno
\cite{spheno}. Once all the calculations described below are included
in NMSSMTools, it can also be used for the MSSM, since the MSSM is just
a particular limiting case of the NMSSM.)

In the remaining part of the introduction we briefly review the
experimental and theoretical status of the various $B$ physics
observables, which are considered in the present paper.

In the past constraints from $b \to s\gamma$ have been particularly
severe, since the experimental world average for $BR(\bar{B} \to X_s
\gamma)$ was somewhat below the (NLO) SM prediction \cite{gm1,hu1},
whereas at least the contribution involving a charged Higgs boson in
the relevant diagram is positive. 

This situation has changed considerably during the last years: the
present world average estimated by the Heavy Flavour Averaging Group
\cite{hfag} reads (for $E_\gamma\; >\; E_0\; =\; 1.6$~GeV)
\beq
\left.BR(\bar{B} \to X_s \gamma)\right|_{exp} = (3.55 \pm
0.24^{+0.09}_{-0.10} \pm 0.03) \times 10^{-4}.
\eeq

The SM NNLO (${\cal O}(\alpha_s^2$)) corrections to the total
$BR(\bar{B} \to X_s \gamma)$ branching fraction have recently be
combined \cite{mis,mist}, which give
\beq\label{1.2e}
\left.BR(\bar{B} \to X_s \gamma)\right|_{SM} = (3.15 \pm 0.23)\times
10^{-4}.
\eeq

In \cite{neubert} the treatment of the cut $E_\gamma\ >\ 1.6$~GeV on
the photon energy has been improved, leading to a still lower SM
prediction:
\beq
\left.BR(\bar{B} \to X_s \gamma)\right|_{SM} = (2.98 \pm 0.26)\times
10^{-4}.
\eeq
This result can be interpreted as (still weak) evidence for BSM
contributions to $b \to s\gamma$; in any case constraints on the
parameter space of Susy models have become considerably less stringent.

Next we turn to $\Delta M_{s,d}$. $\Delta M_s$ has recently been
measured by the CDF collaboration \cite{cdf} with the result
\beq
\Delta M_s^{exp}= 17.77 \pm .12\ ps^{-1}\ .
\eeq
A standard model prediction
\beq
\Delta M_s^{SM} = 20.5 \pm 3.1\ ps^{-1}
\eeq  
can be obtained using a determination of $|V_{ts}^* V_{tb}| = (41.3 \pm
.7)\times 10^{-3}$ from tree level processes (where effects from BSM
physics affect the higher order corrections only) \cite{ball}, and a
determination of $f_{B_s} \sqrt{\hat{B}_{B_s}} = 0.281 \pm .021$ GeV
by the HPQCD collaboration \cite{dalgic}. (In \cite{dalgic}, the
central value $\Delta M_s^{SM} = 20.3\ ps^{-1}$ has been obtained,
since $|V_{ts}^* V_{tb}| = 41.0\times 10^{-3}$ has been used. We note
that here and below the CKM matrix elements are defined in terms of a
low energy effective Lagrangian, whose parameters are determined from
low energy processes \cite{bcrs}. In \cite{bcrs}, these CKM matrix
elements are denoted by $V_{eff}$, but we omit the subscript "${eff}$"
in the following.) Hence, a negative contribution to $\Delta M_s$ from
BSM processes would be welcome.

$\Delta M_d$ is quite well known \cite{hfag}, 
\beq
\Delta M_d^{exp} = 0.507 \pm .004\ ps^{-1}\ .
\eeq
Again, a standard model prediction (see also \cite{freitas})
\beq
\Delta M_d^{SM} =0.59 \pm 0.19\ ps^{-1}
\eeq
can be obtained using a determination of $|V_{td}^* V_{tb}| = (8.6 \pm
1.4)\times 10^{-3}$ from tree level processes \cite{ball}, $f_{B_s}
\sqrt{\hat{B}_{B_s}}$ as above, and $f_{B_s} \sqrt{\hat{B}_{B_s}}/
f_{B_d} \sqrt{\hat{B}_{B_d}} = 1.216 \pm .041$ from \cite {okamoto}.

The various Susy diagrams which contribute to $\Delta M_q$ ($q=s,d$)
are box diagrams involving charged Higgs bosons,  stops and charginos
(see, e.g., \cite{berto}), and double penguin diagrams involving
neutral CP even or CP odd Higgs bosons whose contributions increase
like $\tan^4\beta$ for large $\tan\beta$ (see \cite{bcrs} for a
detailed analysis). As a function of the mass $M_H$ of the Higgs boson,
these contributions to the Wilson coefficients behave like $1/M_H^2$,
and depend on the mixing angles of the CP even and CP odd Higgs mass
matrices. In the MSSM, the dominant contributions $\sim 1/M_h^2$ (where
$h$ denotes the lightest Higgs scalar) cancel
at large $\tan\beta$ \cite{bcrs}, and one is left with contributions
$\sim 1/M_A^2$ (where $A$ denotes the CP odd scalar in the MSSM, whose
mass is close to the heavy CP even scalar for large $M_A$) 
which cannot be too large, given the lower bound on $M_A$ in the MSSM. 

In the NMSSM, three neutral CP even and two CP odd Higgs bosons (we
neglect the Goldstone boson here) contribute to the double penguin
diagrams. Notably the lightest CP odd Higgs boson $A_1$ can be quite
light in the NMSSM and escape the present LEP constraints 
\cite{lighta1,hiller,lighta2,lighta3,lighta4,lighta5,lighta6}, but
with couplings strong enough to generate large effects for low
$M_{A_1}$ \cite{hiller}. Interestingly, the resulting contributions to
$\Delta M_s$ are negative which can improve the agreement with its
measurement.

Neutral Higgs bosons with effective flavour violating couplings
contribute also to\break 
$BR(\bar{B}_s \to \mu^+ \mu^-)$, where the new CDF result is at 95\%
confidence level \cite{cdfmumunew}
\beq 
\left.BR(\bar{B}_s \to \mu^+ \mu^-)\right|_{exp} < 5.8\times 10^{-8}\ .
\eeq 
(At present, constraints from $BR(\bar{B}_d \to \mu^+ \mu^-)$ are less
restrictive.) The SM prediction is still smaller by an order of
magnitude \cite{buchalla,dedes},
\beq 
\left.BR(\bar{B}_s \to \mu^+ \mu^-)\right|_{SM} = (3.8 \pm 0.1)
\times 10^{-9}\ ,
\eeq 
which leaves some room for BSM contributions. Again, a light CP odd
Higgs boson $A_1$ can lead to an important effect in the NMSSM
\cite{hiller}; in the case of $BR(\bar{B}_s \to \mu^+ \mu^-)$,
however, its contribution must not be too large.

Finally we turn to $BR(\bar{B}^+ \to \tau^+ \nu_\tau)$, which has been
observed by the Belle \cite{Belle} and BABAR \cite{babar} experiments.
The actual world average performed by the Heavy Flavor Averaging Group
\cite{hfag} is
\beq\label{1.10e}
\left.BR(\bar{B}^+\rightarrow\tau^+\nu_{\tau})\right|_{exp}=
(1.32 \pm .49) \times 10^{-4}\ .
\eeq

Unfortunately, the corresponding SM prediction is handicapped by a
large uncertainty concerning the CKM matrix element
$\left|V_{ub}\right|$ \cite{hfag,utfit}: Its determination from
inclusive semi\-leptonic $b$ decays gives values near
$\left|V_{ub}\right| \sim 4.4\,\times\,10^{-3}$, whereas its 
determination from exclusive semileptonic decays gives values near
$\left|V_{ub}\right| \sim 3.7\,\times\,10^{-3}$ (leading to a
discrepancy of the order of $2 \sigma$). Accordingly, together with the
uncertainties from the hadronic parameter $f_B$, quite different SM
predictions for $BR(\bar{B}^+ \to \tau^+ \nu_\tau)$ can be obtained,
ranging from
$\left.BR(\bar{B}^+\rightarrow\tau^+\nu_{\tau})\right|_{SM}= (0.85 \pm
.13) \times 10^{-4}$ \cite{carena} to
$\left.BR(\bar{B}^+\rightarrow\tau^+\nu_{\tau})\right|_{SM}= (1.59 \pm
.40) \times 10^{-4}$ \cite{IsP}. Hence we will allow for quite large
theoretical error bars on this process with the result that it hardly
constrains the Susy parameter space; this situation can change in the
future, however.

In section 2 we describe the sources of the contributions to $BR(\bar{B}
\to X_s \gamma)$ that we take into account. In section 3 we give the
sources of our calculations of $\Delta M_s$, $\Delta M_d$,
$BR(\bar{B}_s \to \mu^+ \mu^-)$ and  $BR(\bar{B}^+ \to \tau^+
\nu_\tau)$. In section 4 we present results for $BR(\bar{B} \to X_s
\gamma)$ both for large $\tan\beta$ (relevant for the MSSM and the
NMSSM) and for low $\tan\beta$ (relevant for the NMSSM only). 

Finally, in section 5, we investigate combined constraints from
$BR(\bar{B} \to X_s \gamma)$, $\Delta M_{s,d}$, $BR(\bar{B}_s \to \mu^+
\mu^-)$ and $BR(\bar{B}^+ \to \tau^+ \nu_\tau)$ in parameter regions 
relevant simultaneously for the MSSM and the NMSSM, and also on the
NMSSM specific region involving a light CP odd Higgs scalar. In all
cases we include constraints on the parameter space from LEP on Higgs
masses and couplings as in the updated version of NMHDECAY
\cite{egh,eh3}. In section 6 we conclude with a summary and an outlook.

\section{Computation of $BR(\bar{B} \to X_s \gamma)$} 

The starting point of our computation is the expression for the
branching ratio as in \cite{gm1,hu1},
\bea\label{3.1e}
&BR\left(\bar{B} \to X_s \gamma \right)^{\Psi,
\Psi'\ subtracted}_{E_\gamma > E_0} =&\nn \\
&\left. BR\left(\bar{B} \to X_c e \bar{\nu}\right)\right|_{exp}
\left|\frac{V_{ts}^*V_{tb}}{V_{cb}}\right|^2 \frac{6\alpha_{em}}{\pi
C}&
\left[\left|K_c+r(\mu_0)K_t+\epsilon_{ew}\right|^2+B(E_0)+N(E_0)\right]
\,,
\eea
valid for a matching scale $\mu_0 = m_t(m_t)^{\overline{MS}}$.

In (\ref{3.1e}), we use \cite{mist}
\beq\label{3.2e}
\left. BR\left(\bar{B} \to X_c e \bar{\nu}\right)\right|_{exp}=0.1061\ ,
\eeq
and 
\beq\label{3.3e}
\left|\frac{V_{ts}^*V_{tb}}{V_{cb}}\right|^2 = 0.967
\eeq
from tree level processes \cite{ball}.

$C$ in (\ref{3.1e}) is given by
\beq\label{3.4e}
C = \left|\frac{V_{ub}}{V_{cb}}\right|^2 \frac{\Gamma\left[\bar{B} \to
X_c e \bar{\nu}\right]} {\Gamma\left[\bar{B} \to
X_u e \bar{\nu}\right]}
\eeq
for which we use the numerical value \cite{mist}
\beq\label{3.5e}
C = 0.580\ .
\eeq

$E_0$ is the lower cutoff on the photon energy, for which we chose $E_0
= 1.6$~GeV. $K_t$ includes the SM top quark and the BSM contributions,
whereas $K_c$ denotes the SM charm quark contribution.  $r(\mu_0)$ is
the ratio $m_b^{\overline{MS}}(\mu_0)/m_b^{1s}$, for which we use 
\cite{gm1,hu1}
\beq\label{3.6e}
r(\mu_0)=0.578\left(\frac{\alpha_s\left(M_Z\right)}{0.1185}\right)
\left(\frac{m_b^{1S}}{4.69}\right)^{0.23}
\left(\frac{m_c(m_c)}{1.25}\right)^{-0.003}
\left(\frac{\mu_0}{165}\right)^{-0.08}
\left(\frac{\mu_b}{4.69}\right)^{0.006}
\eeq
with $m_b^{1S} = 4.68$~GeV as in \cite{mist} and $\mu_b=m_b(m_b)$.
(The dependence on the scale $\mu_b$ is in fact negligibly small.)

In (\ref{3.1e}) $\epsilon_{ew}$ denotes the electroweak radiative
corrections, $B(E_0)$ the (gluon) brems\-strahlung corrections, and
$N(E_0)$ are nonperturbative corrections.

Strictly speaking, the expression (\ref{3.1e}) is valid to NLO, where
the charm quark contribution ($K_c$) can be separated from the top
quark/BSM contribution ($K_t$). $K_c$ depends on the ratio $m_c/m_b$,
and hence on the scheme and the scale at which these masses are taken.
On the one hand this ambiguity is a NNLO effect, which is responsible
for the largest part of the theoretical error in the NLO result
\cite{hu1}
\beq\label{3.7e}
\left. BR(\bar{B} \to X_s \gamma)\right|^{NLO} = 
 (\left. 3.61 ^{+0.24}_{-0.40}\right|_{mc/mb} \pm .02_{CKM} 
 \pm 0.24_{param.} \pm 0.14_{scale}) \times 10^{-4}.
\eeq

We found that the NNLO result (\ref{1.2e}) is reproduced (for
$m_{t,pole}=171.4$~GeV, as assumed in \cite{mis,mist}), if one uses the
relatively large value
\beq\label{3.8e}
\frac{m_c}{m_b} = 0.307
\eeq
(close to the pole quark masses) in the expression for $K_c$. We
believe that as long as the BSM contributions -- which are added
linearly to the SM contributions in the factor $K_t$ -- are not
evaluated to NNLO, the error arising from this procedure is not larger
than the error intrinsic to the BSM contributions (which is estimated
quite conservatively below). It is guaran\-teed, in any
case, that the result for the $BR(\bar{B} \to X_s \gamma)$ in the
decoupling limit of the BSM contributions assumes the NNLO SM value in
(\ref{1.2e}).

Subsequently we describe the origin of the formulas used for our
evaluation of the quantities $K_c$, $N(E_0)$, $B(E_0)$, $\epsilon_{ew}$
and $K_t$ in (\ref{3.1e}). First, $K_c$ is computed as in Eq. (3.7) in
\cite{gm1}, with $\mu_b = m_b$, the value (\ref{3.8e}) for $m_c/m_b$
and
\beq\label{3.9e}
\mu_0 = m_t(m_t)^{\overline{MS}}
\eeq
for the matching scale $\mu_0$. The ratio of CKM matrix elements
$\epsilon_s$, that appears in Eq. (3.7) in \cite{gm1}, is taken from
\cite{mist}:
\beq\label{3.10e}
\epsilon_s\equiv V^*_{us}V_{ub}/(V^*_{ts}V_{tb}) = -0.011 + i\ 0.0180
\eeq 

The nonperturbative corrections $N(E_0)$ are computed as in Eq. (3.10)
in \cite{gm1} in terms of the lowest order coefficients $K_c^{(0)}$ and
$K_t^{(0)}$ (including the BSM contributions to the latter), with
$\lambda_2 = 0.12$~GeV$^2$. ($N(E_0)$ is actually independent from
$E_0$ in this approximation).

The bremsstrahlung corrections $B(E_0)$ are taken from the appendix E
in \cite{gm1} with, we repeat, an energy cutoff $E_0 = 1.6$~GeV. For
the parameter $z = (m_c/m_b)^2$ we use a value consistent with eq.
(\ref{3.8e}) above. (In any case the dependence of $B(E_0)$ on $z$ is
weak \cite{gm1}.) The corrections $\sim \epsilon_q$ (with $q=s$) as in
Eq. (28) in \cite{hu1} are taken into account, with $\epsilon_s$ given
in Eq. (\ref{3.10e}) above. The contributions to $B(E_0)$ from the
coefficients $C_i^{(0)}$ with $i=3\dots 6$ are neglected as in
\cite{gm1}, on the other hand the BSM contributions to the coefficients
$C_7^{(0)}$ and $C_8^{(0)}$ are taken into account.

For the electroweak corrections $\epsilon_{ew}$ in (\ref{3.1e}) we use
the formula (3.9) in \cite{gm1} (see also Eq. (27) in \cite{hu1}),
which gives a SM contribution $\epsilon_{ew}^{SM}=0.0071$ according to
\cite{gh}. To this SM value for $\epsilon_{ew}$ we add the BSM
contributions as in \cite{gm1,hu1} in terms of the BSM contributions to
the coefficients $C_{7,8}$ discussed below.

Finally we turn to the calculation of $K_t$ including the BSM
contributions. First, the SM contributions to $K_t$ (including the NLO
in $\alpha_s$) are taken from Eq. (3.8) in \cite{gm1}, with the above
Eq. (\ref{3.9e}) for the matching scale $\mu_0$. The BSM contributions
are added as in Eq. (5.1) in \cite{gm1}. The BSM contributions appear
in the LO Wilson coefficients $C_7^{(0) BSM}(\mu_0)$, $C_8^{(0)
BSM}(\mu_0)$ and the NLO Wilson coefficients $C_7^{(1) BSM}(\mu_0)$,
$C_8^{(1) BSM}(\mu_0)$ and $C_4^{(1) BSM}(\mu_0)$ of the corresponding
operators $P_i$.

Our calculation of these Wilson coefficients within the MSSM and the
NMSSM starts with the calculation of the corrections $\epsilon_b$,
$\epsilon_b'$ and $\epsilon_t'$ to the couplings of the charged Higgs
bosons to quarks defined in \cite{dgg} (see also \cite{car}), which are
important at large $\tan\beta$. We use the expressions for these
parameters given in  \cite{micro2}, which include sbottom and
electroweak contributions, and in which a sign error in \cite{dgg} is
corrected. In the case of $\epsilon_b'$ and $\epsilon_t'$ we sum over
the 5 neutralino states of the NMSSM with its corresponding masses and
couplings. (In the MSSM limit $\lambda$, $\kappa \to 0$ of the NMSSM,
the fifth neutralino decouples and does not contribute.) Then we proceed
with the computation of the following BSM contributions to the Wilson
coefficients:

a) The chargino-squark loop contributions to $C_7^{(0)}$ and
$C_8^{(0)}$ (as, e.g., in appendix B in 
\cite{micro2}), computed again at $M_{Susy}$ and evolved to our
matching scale $\mu_0$. Corresponding NLO corrections are known in the
particular case where one stop is lighter than the other squarks and
the gluino \cite{cdgg1}, and the complete QCD corrections have been
computed in \cite{dgs}, but here we content ourselves with the
summation of the leading logarithms of the ratio $M_{Susy}/\mu_0$ via
the RG evolution of the Wilson coefficients.

b) The charged Higgs--top-quark loop contributions to $C_7^{(0)}$ and
$C_8^{(0)}$ (again as in appendix B in 
\cite{micro2}), and the corresponding NLO contributions to
$C_4^{(1)}$, $C_7^{(1)}$ and $C_8^{(1)}$ \cite{cdgg2}. The LO
contributions to $C_7^{(0)}$ and $C_8^{(0)}$ are evolved from the scale
corresponding to the charged Higgs mass to our matching scale $\mu_0$,
and we took care not to include large logarithms -- that appear
potentially also in the NLO contributions -- twice. (Higher order large
$\tan\beta$ corrections to the NLO contributions are neglected.)

c) As in \cite{agis} we take the neutral Higgs contributions to the
Wilson coefficients $C_7^{(0)}$ and $C_8^{(0)}$ into account following
Eq. (6.61) in \cite{bcrs}. However, contrary to $\Delta M_q$ and
$BR(\bar{B}_s\rightarrow \mu^+\mu^-)$ below, these neutral Higgs
effects remain small and usually inside our theoretical error bars. 

d) Finally the large $\tan\beta$ corrections induce also a shift in the
SM contributions to the coefficients $C_7^{(0)}(\mu_0)$ and
$C_8^{(0)}(\mu_0)$ \cite{cdgg2,micro2}.

Herewith we have described completely the origins of the 
considered contributions to $BR(\bar{B} \to X_s \gamma)$.

\section{$\Delta M_{q}$, $BR(\bar{B}_s \to \mu^+ \mu^-)$ and
$BR(\bar{B}^+ \to \tau^+ \nu_\tau)$}

In this section, we discuss the sources for our evaluation of the $B$
physics observables $\Delta M_q$ ($q=s,d$),
$BR(\bar{B}_s\rightarrow\mu^+\mu^-)$ and
$BR(\bar{B}^+\rightarrow\tau^+\nu_{\tau})$. The formula for $\Delta
M_q$ is taken from \cite{bcrs}, eqs. (6.6--7):
\beq
\Delta M_q = \frac{G_F^2M_W^2}{6\pi^2} M_{B_q} \eta_B
f_{B_q}^2\hat{B}_{B_q}
\left|V_{tq}^{*}V_{tb}\right|^2
\left|F^q_{tt}\right|
\eeq
with
\beq
F_{tt}^q = S_0(x_t)+\frac{1}{4r}C_{new}^{VLL}  
+\bar{P}_1^{SLL}\left(C_1^{SLL}+C_1^{SRR}\right)
+\bar{P}_2^{LR}C_2^{LR}
+ \dots
\eeq
where we have omitted neglicibly small contributions, and where we
take \cite{bcrs} $r=0.985$, $\bar{P}_1^{SLL}=-0.37$, 
$\bar{P}_2^{LR}=0.90$ and $\eta_B=0.551$. We use the meson masses
$M_{B_d}=5.2794$~GeV and $M_{B_s}=5.3675$~GeV, and the hadronic
parameters $f_{B_s}\sqrt{\hat{B}_{B_s}}=0.281$~GeV from \cite{dalgic}
and $f_{B_d}\sqrt{\hat{B}_{B_d}}=0.231$~GeV from $f_{B_s}
\sqrt{\hat{B}_{B_s}}/ f_{B_d} \sqrt{\hat{B}_{B_d}} = 1.216$
\cite{okamoto}.  As stated in the introduction, we use the CKM factors
deduced from tree level processes, which are less sensitive to BSM
physics:  $|V_{ts}^{*} V_{tb}| = 41.3 \times 10^{-3}$ and  $|V_{td}^{*}
V_{tb}| = 8.6 \times 10^{-3}$ \cite{ball}. $S_0$ in $F_{tt}$ stands for
the SM contribution ($x_t\equiv
\left(\frac{m_t^{\overline{MS}}}{M_W}\right)^2$), whereas  the
coefficients $C_{1,2}^{\, i}$ contain BSM contributions to the
corresponding effective 4-quark operators.

Let us discuss the various contributions to $F^q_{tt}$
which we take into account (we repeat that we assume  minimal flavor
violation such that the only source of flavor violation is the CKM
matrix):  The SM contribution originates from quark/$W^{\pm}$ box
diagrams. In multi-Higgs extensions of the SM such as the MSSM or the
NMSSM, charged Higgs bosons can replace one or both $W^{\pm}$ bosons in
these box diagrams. A second type of box diagrams arises in Susy from
squark/chargino loops. All these box contributions are calculated as in
eqs. (93--95) in \cite{berto} and added directly to $S_0$:
\beq
S_0\rightarrow
S_0+x_t\left(\Delta_{H^{\pm}}+\Delta^q_{\chi_{\pm}}\right)
\end{equation}

We have checked that at low $\tan\beta$, where the box contributions
are most significant, the results in \cite{alt} are reproduced.

Double Penguin diagrams involving a neutral Higgs propagator connecting
two flavor changing effective vertices can be significantly enhanced
for large $\tan\beta$ or light scalars. We closely follow the analysis
carried out in \cite{bcrs}:
\begin{itemize}
\item First, we compute flavor dependent $\varepsilon$ parameters
(effective vertices) arising from loops involving sparticles in the
effective Lagrangian describing the Higgs quark couplings. We use Eq.
(5.1) and appendix A.2 in \cite{bcrs}. However, we extend the
neutralino sector according to the NMSSM; the corresponding
generalization of the MSSM formulae is straightforward.
\item Next, we define flavor-changing neutral Higgs-quark couplings
$X_{LR/RL}^{S\,bs}$ as in Eqs. (3.55--56) in \cite{bcrs} ($S$ denote the
various neutral Higgs bosons).
The corresponding Higgs mixing angles $x_d^S$ and $x_u^S$ can be 
generalized in a straightforward way to the NMSSM using the
decomposition of the neutral weak eigenstates  $H_u^{0}$ and $H_d^{0}$
into the neutral physical states $S^0$ (in the convention of
\cite{bcrs}) as $H_u^{0\,*}=\sum_{S^0} x_u^SS^0$, $H_d^{0}=\sum_{S^0}
x_d^SS^0$.
\item Finally, we use Eq. (6.12) of \cite{bcrs} (neglecting the
Goldstone boson contribution) for the three relevant coefficients
$C_1^{SLL}$, $C_1^{SRR}$ and $C^{LR}_2$. However, as we will face very
light (pseudo)scalar masses (possibly below $10$~GeV in some parts of
the NMSSM parameter space), we can no longer be content with the
approximation $\frac{1}{m_S^2}$ for the scalar propagator (see
\cite{hiller}, Eq. (32)). Thus, we replace these factors by
Breit-Wigner functions:
\begin{equation}
\frac{1}{m_S^2}\rightarrow
\frac{sgn(m_S^2-M_{B_q}^2)}{\sqrt{\left(m_S^2-M_{B_q}^2\right)^2+
m_S^2\Gamma_S^2}}
\end{equation}
\end{itemize}
(The width $\Gamma_S$ is computed as in NMSSMTools \cite{egh,eh3}.)
In the MSSM, relations between the Higgs masses at large $\tan\beta$
allow for further simplifications of the final formula for $\Delta M_q$
(see \cite{bcrs}, Eq. (6.23)). However, in the NMSSM a correct
description of the singlet like contributions does not allow for such
simplifications.

Next we consider $BR(\bar{B}_s\rightarrow \mu^+\mu^-)$. We calculate 
the Branching Ratio according to Eq. (5.15-16) of 
\cite{Bobeth} (we neglect the $c'_i$):
\bea
&&BR(\bar{B}_s\rightarrow\mu^+\mu^-)=\nn\\
&&\frac{G_F^2 \alpha^2 M_{B_s}^5 f_{B_s}^2 \tau_{B_s}}
{64\pi^3\sin^4 \theta_W}\left|V_{tb}V_{ts}^*\right|^2
\sqrt{1-4\frac{m_{\mu}^2}{m_{B_s}^2}}
\left[\frac{1-4\frac{m_{\mu}^2}{M_{B_s}^2}}
{\left(1+\frac{m_s}{m_b}\right)^2}\left|c_S\right|^2
+\left|\frac{c_P}{1+\frac{m_s}{m_b}}+\frac{2m_{\mu}}{M_{B_s}^2}
c_A\right|^2\right]
\eea
where $c_A$ contains the SM contribution arising from box and penguin
diagrams, which is one order of magnitude below the sensitivity of
present experimental data. The neutral Higgs contributions to $c_S$ and
$c_P$ are the only ones which could lead to a significant deviation
from the SM prediction. The corresponding diagrams involve the
effective flavour violating neutral Higgs vertex and a neutral Higgs
propagator. We infer from an appropriate generalization of the
equations given in \cite{bcrs} the appropriate formulae for the
coefficients $c_S$ and $c_P$ in the NMSSM. Again, it proves necessary
to replace the approximation $\frac{1}{m_S^2}$ by a Breit-Wigner
function.

Charged Higgs corrections to $BR(\bar{B}^+\rightarrow\tau^+\nu_{\tau})$
were studied in \cite{Hou} and lead to a destructive interference with
the SM ($W^{+}$) contribution:
\begin{equation}
BR(\bar{B}^+\rightarrow\tau^+\nu_{\tau})=\frac{G_F^2M_Bm_{\tau}^2}{8\pi}
\left(1-\frac{m_{\tau}^2}{M_B^2}\right)^2f_B^2
\left|V_{ub}\right|^2\tau_B\,r_H\ ,
\end{equation}
where $r_H$ parametrizes the deviation from the SM prediction. The
expression for $r_H$ has been improved in \cite{Akeroyd} in order to
take large $\tan\beta$ corrections into account:
\begin{equation}
r_H=\left[1-\left(\frac{M_B}{m_{H^{\pm}}}\right)^2
\frac{\tan^2\beta}{1+\tilde{\varepsilon}_0\tan\beta}\right]^2
\end{equation}

Having described the origin of all relevant calculations, we turn to
the numerical results, concentrating first on $BR(\bar{B} \to X_s
\gamma)$.

\section{Results for $BR(\bar{B} \to X_s \gamma)$ in the MSSM and
the NMSSM} 

The BSM contributions to $\bar{B} \to X_s \gamma$ depend
essentially on the charged Higgs mass, $\tan\beta$ and, for large
$\tan\beta$, on $A_t$.

\begin{figure}[t]
\begin{center}
\rotatebox{-90}{\epsfig{file=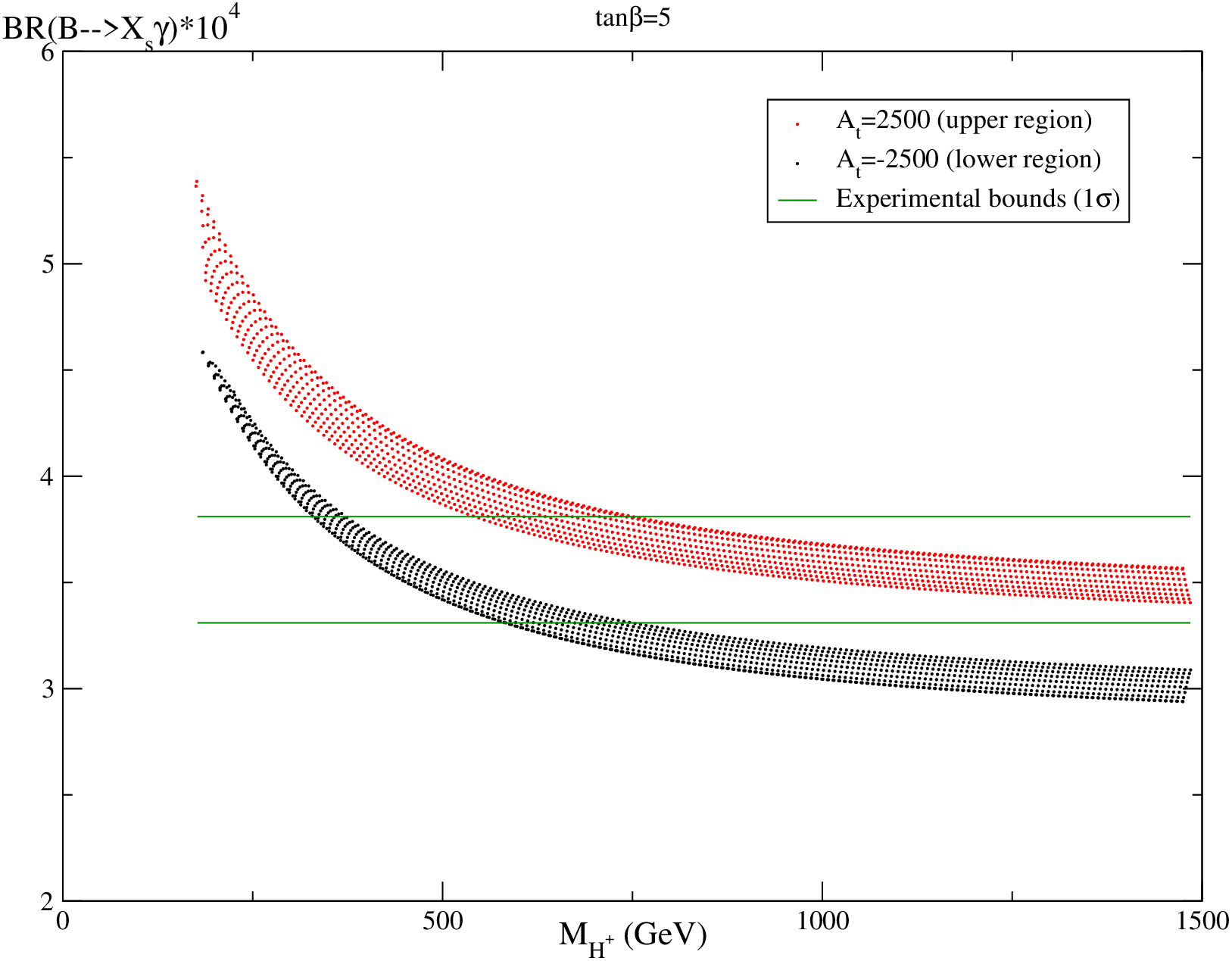,height=15cm}}
\end{center}
\caption{$BR(B\rightarrow X_s \gamma)$ as a function of the charged
Higgs mass, for $\tan\beta=5$, $A_t=\pm2500$~GeV. The green lines
represent the experimental $1\sigma$ bounds.}
\end{figure}

First, we focus on the impact of the charged Higgs mass on $BR(\bar{B}
\to X_s \gamma)$, which is always positive. The branching ratio is a
decreasing function of $m_{H^{\pm}}$, since the contributions from
charged Higgs diagrams decay like $1/m_{H^{\pm}}^4$.
Before the recent improvements on the experimental side and the SM
contributions discussed in the introduction, quite severe bounds on
$m_{H^{\pm}}$ could be deduced notably for small to modest values of
$\tan\beta$, where the additional Susy contributions (which can have
both signs, depending on the relative sign of $A_t$ to $\mu$)
cannot be too large in absolute value.

The updated situation is described in Figs. 1--4. In Fig.~1 we show our
results for $\bar{B} \to X_s \gamma$ for $\tan\beta =5$, universal
squark masses of 1~TeV, gaugino masses $M_1 = 150$~GeV, $M_2 =
300$~GeV,  $M_3 = 900$~GeV, for two extreme values of $A_t =
+2.5$~TeV and -2.5~TeV as a function of $m_{H^{\pm}}$. We scan over the
parameter  $\mu$ between +100~GeV and +1~TeV, which explains the
broadening of the two dotted distributions. (The inner regions
correspond to larger values of $\mu$, the outer regions to the lowest
value of $\mu$ that is allowed by the non-observation of charginos.)
For the top quark mass we take 171.4~GeV.
The 1$\sigma$ experimentally allowed region is also indicated and it
becomes clear that, at least after taking theoretical errors into
account (see below), relatively low values of $m_{H^{\pm}}$ down to
$\sim 200$~GeV cannot be excluded. This result holds both for the MSSM
and the NMSSM (where the $\mu$-parameter has to be replaced by an
effective parameter $\mu_{eff} = \lambda \left< S \right>$, we use the
conventions of \cite{egh}); no dependence on the additional parameters
of the NMSSM remains visible.

\begin{figure}[t]
\begin{center}
\rotatebox{-90}{\epsfig{file=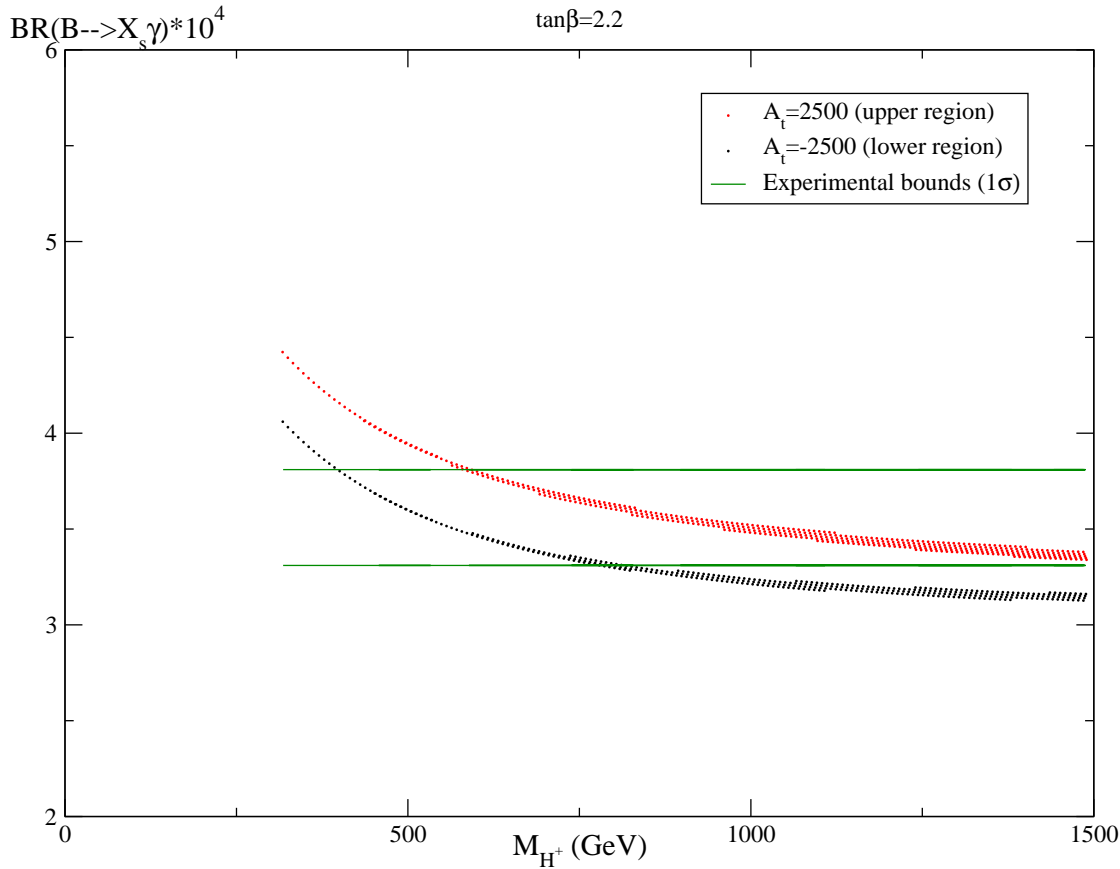,height=15cm}}
\end{center}
\caption{$BR(B\rightarrow X_s \gamma)$ as a function of the charged
Higgs mass, for $\tan\beta=2.2$, $A_t=\pm 2500$~GeV. The green lines
represent the experimental $1\sigma$ bounds.}
\end{figure}

\begin{figure}[ht]
\begin{center}
\rotatebox{-90}{\epsfig{file=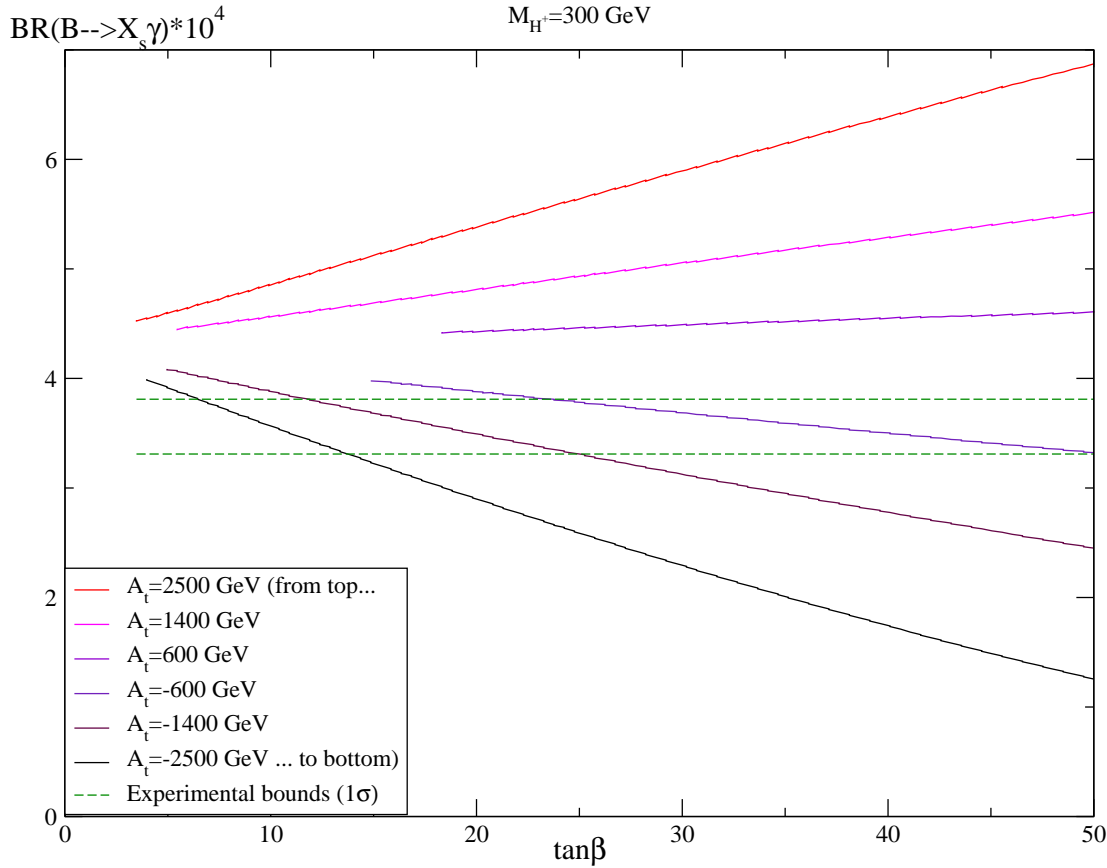,height=15cm}}
\end{center}
\caption{$BR(B\rightarrow X_s \gamma)$ as a function of $\tan\beta$,
for $M_{H^+}=300$~GeV and various values of $A_t$}
\end{figure}

Before we turn to larger values of $\tan\beta$, we study a relatively
low value $\tan\beta = 2.2$ which would make it very difficult for the
MSSM to satisfy the constraints from LEP on the lightest neutral Higgs
boson mass, but which is perfectly consistent in the NMSSM \cite{eh1},
even in the CNMSSM with universal soft terms at the GUT scale
\cite{eh2}. Fig. 2 is the same as Fig. 1, but for $\tan\beta = 2.2$ and
NMSSM parameters $\lambda = 0.5$, $\kappa = 0.4$ and $A_\kappa = -200$
GeV, which lead to neutral Higgs masses consistent with LEP
bounds provided $m_{H^{\pm}} \gsim 300$~GeV (due to correlations
between the various Higgs mass matrices in the NMSSM). There is no
particular impact of the NMSSM parameters on $\bar{B} \to X_s \gamma$, 
however. One finds that this NMSSM specific region in
parameter space is hardly constrained by this observable.

Next we investigate $\bar{B} \to X_s \gamma$ for
larger values of $\tan\beta$. We find an approximate linear dependence
on $\tan\beta$ with a slope determined essentially by $A_t$, at
least for given $\mu$, which we fix now at 300~GeV. In Fig. 3 we show
our results for various values of $A_t$, $m_{H^{\pm}}=300$~GeV (and
the same other parameters as above), and in Fig. 4 for
$m_{H^{\pm}}=1$~TeV (which is obtained essentially by a vertical shift
of Fig. 3). Now one finds that, the larger $\tan\beta$, the stronger
are constraints on $\left|A_t\right|$ from $BR(\bar{B} \to X_s
\gamma)$. On the other hand, positive contributions from relatively
light charged Higgses can easily be cancelled by appropriate
contributions from squark/chargino loops. Again, these results hold both
for the MSSM and the NMSSM. We note that all points/lines shown in
our Figures correspond to parameters which satisfy LEP constraints on
Susy Higgs bosons, but this is not always trivial: small values of
$\left|A_t\right|$ and $\tan\beta$ can lead to a too light neutral
Higgs boson both in the MSSM and in the NMSSM; this is the reason why
we confined ourselves to $\left|A_t\right| \geq 600$~GeV in Figs. 3 and
4, and why the lines (notably for $\left|A_t\right| = 600$~GeV) do not
continue to arbitrarily small values of $\tan\beta$.

\begin{figure}[t]
\begin{center}
\rotatebox{-90}{\epsfig{file=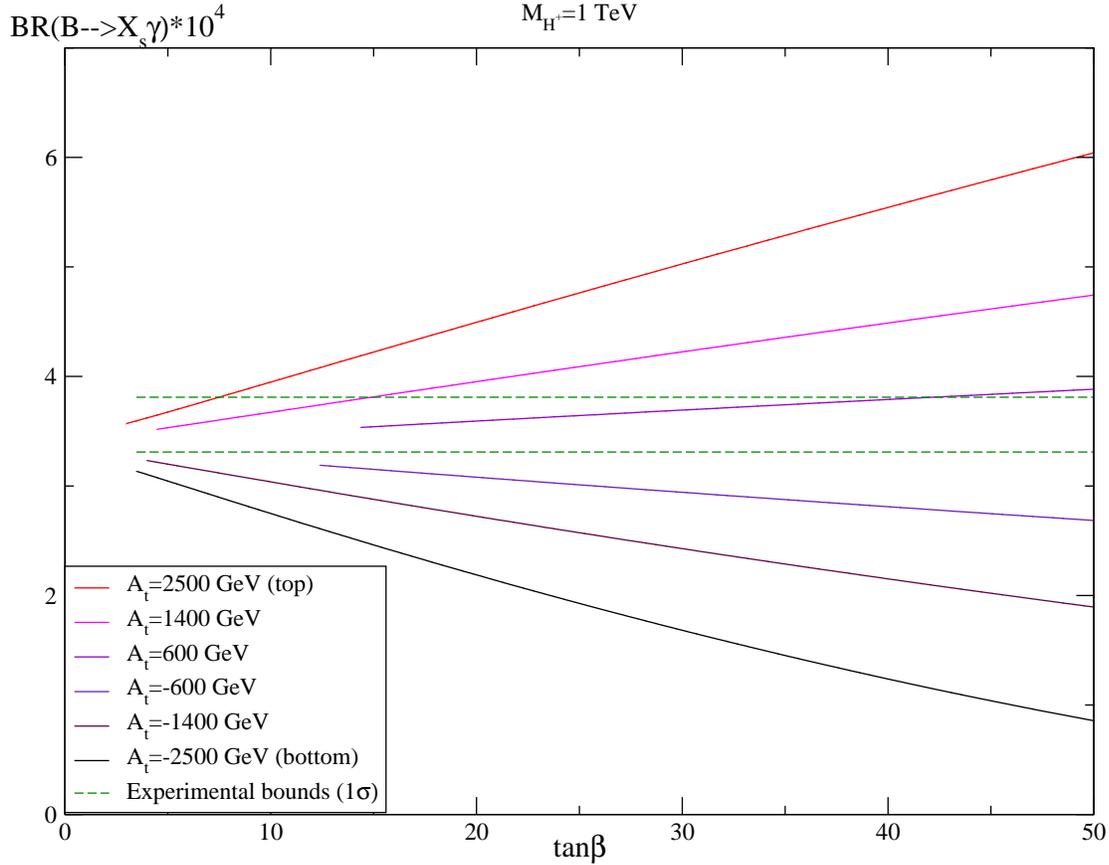,height=15cm}}
\end{center}
\caption{$BR(B\rightarrow X_s \gamma)$ as a function of $\tan\beta$,
for $M_{H^+}=1$~TeV and various values of $A_t$.}
\end{figure}

Although further dependencies on, e.g., the soft Susy breaking squark
and gaugino masses would certainly merit further studies (which can be
performed using NMSSMTools \cite{nmtools}, once
updated), we believe that our Figs. 1--4 represent a fairly
comprehensive review of the actual status of the predictions for
$BR(\bar{B} \to X_s \gamma)$ in the MSSM and the NMSSM.

\section{Constraints from $BR(\bar{B} \to X_s \gamma)$,  $BR(\bar{B}_s
\to \mu^+ \mu^-),$\newline $\Delta M_{q}$, and $BR(\bar{B}^+ \to
\tau^+ \nu_\tau)$ in the MSSM and the NMSSM} 

The aim of this chapter is to study the combination and the relative
relevance of the constraints on the parameter space of the MSSM and the
NMSSM from $BR(\bar{B} \to X_s \gamma)$, $BR(\bar{B}_s \to \mu^+
\mu^-),$ $\Delta M_{s,d}$ and $BR(\bar{B}^+ \to \tau^+ \nu_\tau)$. 
To this end we need to estimate the theoretical error implicit in our
calculations. We intend to remain conservative and to denote a point as
excluded only if one of the observables falls outside the 95\%
confidence limit (or~2$\sigma$). 

In the case of $BR(\bar{B} \to X_s \gamma)$, the theoretical error will
depend on the parameters of the Susy model under consideration; a
general value for the theoretical error would be misleading. Hence we
estimate the theoretical error separately for the charged Higgs, Susy
and SM contributions to $BR(\bar{B} \to X_s \gamma)$ as follows: Since
the charged Higgs contribution is evaluated to NLO, we assume that its
relative theoretical error is only 10\%. For the Susy contribution,
which is evaluated to LO only (up to leading logarithms), we assume a
(conservative) relative theoretical error of 30\%. Finally we estimate
the theoretical error bars of the SM contribution as follows: Given
that the improved treatment of the cut on the photon energy in
\cite{neubert} leads to a lower SM prediction than in \cite{mis}, we
allow the SM contribution to $BR(\bar{B} \to X_s \gamma)$ to vary in
the range $2.72\times 10^{-4}$ to $3.38\times 10^{-4}$. The SM and BSM
errors are added linearly, which gives our estimate of the final
theoretical error.

For $\Delta M_q$ and $BR(\bar{B}_s\rightarrow\mu^+\mu^-)$, we estimate
the theoretical error due to BSM contributions to be of the order of
$30\%$, since no QCD corrections are taken into account. We add these
uncertainties linearly to the $2\sigma$ SM error bars, which gives our
complete theoretical error estimate. (The $1\sigma$ SM error bars on
$\Delta M_q$, arising mostly from the uncertainties of CKM matrix
elements and lattice computations of hadronic parameters, are given in
eqs. (1.5) and (1.7) above.) 

Concerning $BR(\bar{B}^+\rightarrow\tau^+\nu_{\tau})$, the
uncertainties originating from the CKM matrix element
$\left|V_{ub}\right|$ are considerable. We allow $\left|V_{ub}\right|$
to vary in the range $3.3\times 10^{-3} \lsim \left|V_{ub}\right| \lsim
4.7 \times 10^{-3}$, with $4.0\times 10^{-3}$ as central value. For
$f_B$ we use $f_B=0.216 \pm 0.022$ GeV as obtained by the HPQCD
collaboration \cite{hpqcd}. It just so happens that the corresponding
central values lead to a SM prediction
$\left.BR(\bar{B}^+\rightarrow\tau^+\nu_{\tau})\right|_{SM} =
1.32\times 10^{-4}$ in agreement with the experimental central value
given in (\ref{1.10e}). Allowing for $2 \sigma$ error bars on $f_B$ and
the experimental average (\ref{1.10e}), and using the above range for
$\left|V_{ub}\right|$ one finds that $r_H$ in Eq. (3.7), neglecting
additional theoretical errors, is allowed to vary over the quite large
range
\beq
0.13 \lsim r_H \lsim 4.0\ .
\eeq
Consequently the constraints on the parameters $\tan\beta$ and
$m_{H^\pm}$ from this process are typically less stringent than the
ones from other processes.

Now we turn to the dependence of the observables on the most relevant
parameters.

$\Delta M_q$ and $BR(\bar{B}_s\rightarrow \mu^+\mu^-)$ are quite
sensitive to (double) Penguin contributions involving neutral Higgs
bosons. These contributions are controlled by the parameter
$\left(\frac{x_d^S\tan^{\nu}\beta}{m_S^2}\right)^2$, where $\nu=3$ for
$BR(\bar{B}_s\rightarrow \mu^+\mu^-)$ and $\nu=2$ for $\Delta M_q$ (and
$x_d^S$ denotes the $H_d$ component of the neutral Higgs boson $S$);
this explains why $BR(\bar{B}_s\rightarrow \mu^+\mu^-)$ is usually
more sensitive to neutral Higgs effects, at least at large $\tan\beta$
where they can become huge, leading to a violation of experimental
bounds both in the MSSM and the NMSSM.

Thus, in general, large values of $\tan\beta$ are rather strongly
constrained by these observables. However, it is still possible to
reduce the neutral Higgs contributions by assuming heavy scalars
and pseudoscalars (through a large doublet mass $M_A\sim M_{H^{\pm}}$).
Another way to circumvent these constraints consists in assuming
parameters as the trilinear soft-coupling $A_t$ or $\mu_{eff}$ such
that the $\varepsilon$ parameters (which control the flavour violating
neutral Higgs couplings) remain small enough -- here cancellations are
often possible.

Only for low $\tan\beta$ can the positive contributions from Susy box
diagrams to $\Delta M_s$ be more important than the double Penguin
contributions. Given that the SM prediction for $\Delta M_s$
\cite{dalgic} is already $\sim 1\sigma$ above the CDF result
\cite{cdf}, such additional positive BSM contributions could in
principle exclude points in the parameter space at low $\tan\beta$.
(For larger $\tan\beta$ the double Penguin diagram, which gives a
negative contribution $\Delta M_q$, usually dominates the box
diagrams.) However, once we use $2\sigma$ error bars for the CKM matrix
element and hadronic uncertainties, such exclusions at low $\tan\beta$
occur scarcely in practice.

In the following we present several examples of constraints on
$\tan\beta$ and $M_{H^{\pm}}$ (for fixed other parameters)
that originate from the above processes.

First we consider the MSSM and the NMSSM for relatively small values of
$\lambda$ and $\kappa$ ($\lsim$ 0.1), for which the contributions to the
above processes are practically the same in both models. For the soft
Susy breaking squark and gaugino masses we take the same values as in
Figs. 1--4, and 300~GeV for $\mu$ (or $\mu_{eff}$ in the NMSSM).

In Fig. 5 we assume $A_t = 2.5$~TeV.  Dark dotted regions are
excluded by LEP: Here and in Figs. 6 and 7 below the non-observation of
a light neutral Higgs scalar $h$ at LEP implies lower limits on the
MSSM parameter $M_A$ (depending on $\tan\beta$ and $A_t$) which, in
turn, lead to lower limits on $M_{H^\pm}$ ($\sim M_A$ for large $M_A$).
The domain allowed by LEP is further constrained by B physics
processes. We note that the $BR(\bar{B} \to X_s \gamma)$ is by far the
most stringent constraint in Fig.~5. It is indeed particularly
severe because both the charged Higgs and the Susy contributions are
positive and thus cannot balance each other. (Constraints from  $\Delta
M_{d}$ are never more restrictive than constraints from $\Delta M_{s}$,
hence $\Delta M_{q}$ in Figs. 5 and 6 stands for $\Delta M_{s}$.)

\begin{figure}[ht]
\begin{center}
\rotatebox{-90}{\epsfig{file=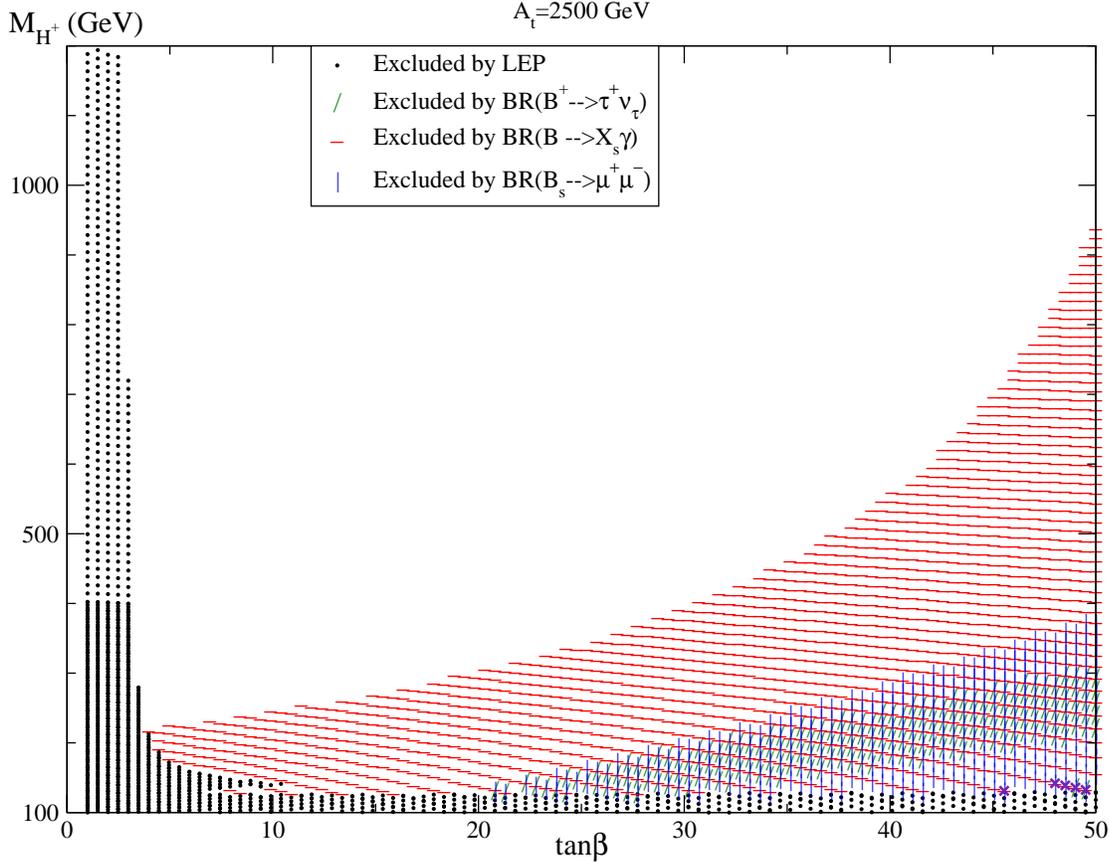,height=15cm}}
\end{center}
\caption{Constraints in the $\tan\beta$-$M_{H^+}$ plane for
$A_t=2500$~GeV.}
\end{figure}

\begin{figure}[t]
\begin{center}
\rotatebox{-90}{\epsfig{file=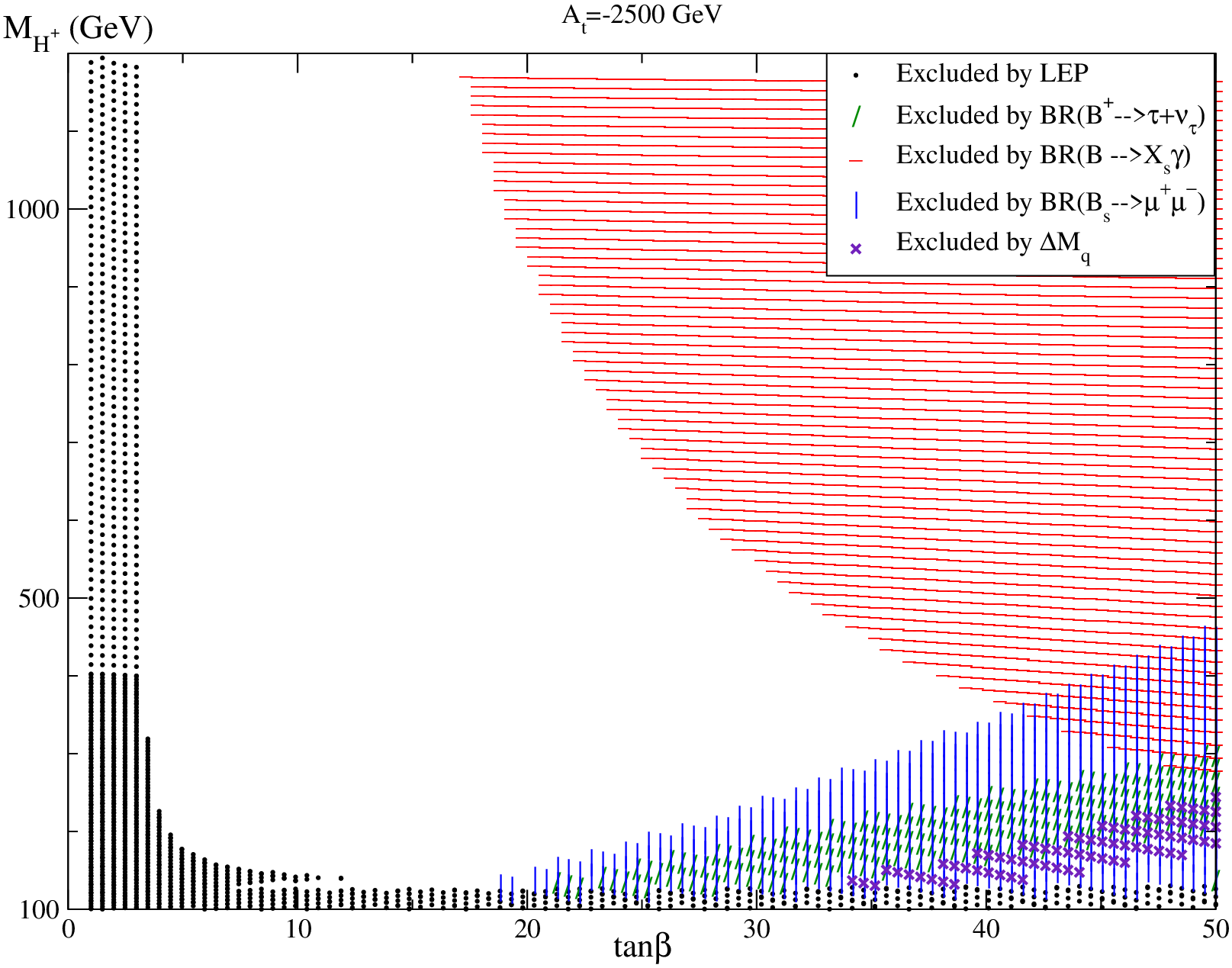,height=15cm}}
\end{center}
\caption{Constraints in the $\tan\beta$-$M_{H^+}$ plane for
$A_t=-2500$~GeV.}
\end{figure}

In Fig. 6 we switch to $A_t = -2.5$~TeV, where the situation is
quite different (the notation is the same as in Fig. 5):
$BR(\bar{B} \to X_s \gamma)$ allows for additional domains, which 
originate from  cancellations between the charged Higgs and Susy
contributions (strongly enhanced by the large value of
$\left|A_t\right|$). Therefore, light charged Higgs bosons (with masses
down to $\sim 100$~GeV) are not excluded by this process; on the
contrary, for $\tan\beta \gsim 20$, they must be light enough to avoid
a large decrease of the branching ratio due to Susy diagrams. However,
these regions are also constrained by  $BR(\bar{B}_s\to \mu^+ \mu^-)$ 
and, less stringently, by $BR(\bar{B}^+ \to \tau^+ \nu_\tau)$ and
$\Delta M_{q}$.

\begin{figure}[t]
\begin{center}
\rotatebox{-90}{\epsfig{file=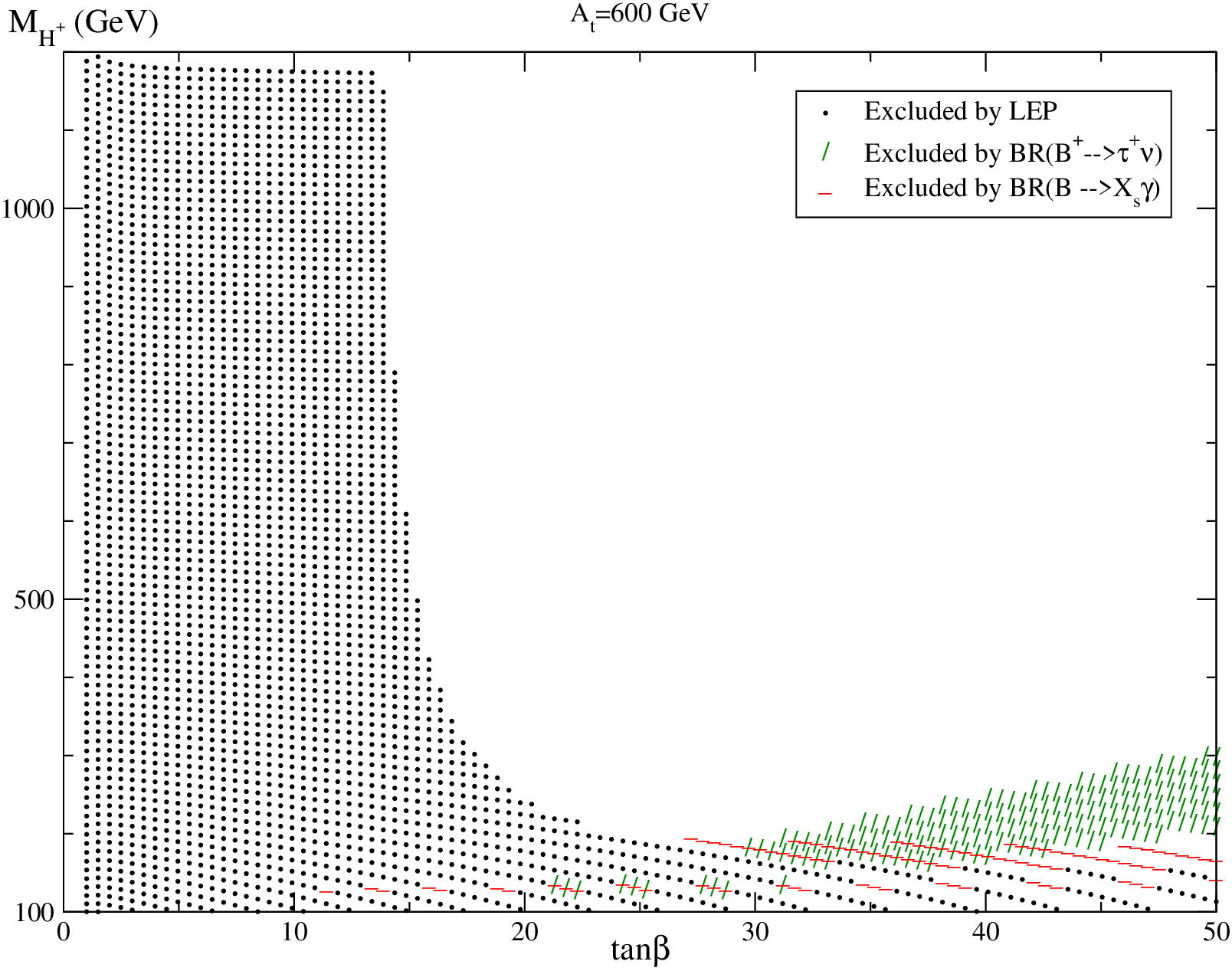,height=15cm}}
\end{center}
\caption{Constraints in the $\tan\beta$-$M_{H^+}$ plane for
$A_t=600$~GeV.}
\end{figure}

In Fig. 7 we consider smaller values of $\left| A_t\right|$, $A_t =
600$ GeV: Now, small values of $\tan\beta$ and $M_{H^\pm}$ (or $M_A$)
are ruled out by LEP constraints on $M_h$. (The precise bound is very
sensitive to radiative corrections to $M_h$ and hence to $m_{top}$. We
recall that we use $m_{top} = 171.4$~GeV.) LEP constraints do not rule
out a narrow strip around $M_{H^\pm}\sim 120$~GeV (already visible in
Fig. 6), where the coupling of $h$ to the Z-Boson is suppressed (since
the MSSM-like parameter $\sin(\beta - \alpha)$ happens to be small) and
where $M_h\sim 100$~GeV.  However, even this region is now excluded by
the charged Higgs contribution to $BR(\bar{B} \to X_s \gamma)$. (For
positive or small absolute values of $A_t$ the Susy contribution to
$BR(\bar{B} \to X_s \gamma)$ cannot cancel the charged Higgs
contribution.) $BR(\bar{B}_s\to \mu^+ \mu^-)$ does no longer lead to
constraints since neutral Higgs effects, which are (roughly)
proportional to $A_t$, remain small for a low value of this parameter.
On the contrary, $BR(\bar{B}^+ \to \tau^+ \nu_\tau)$, which depends
only weakly on $A_t$, can become the dominant B physics constraint for
$\tan\beta \gsim 30$.

\begin{figure}[t]
\begin{center}
\includegraphics[scale=0.55,angle=-90]{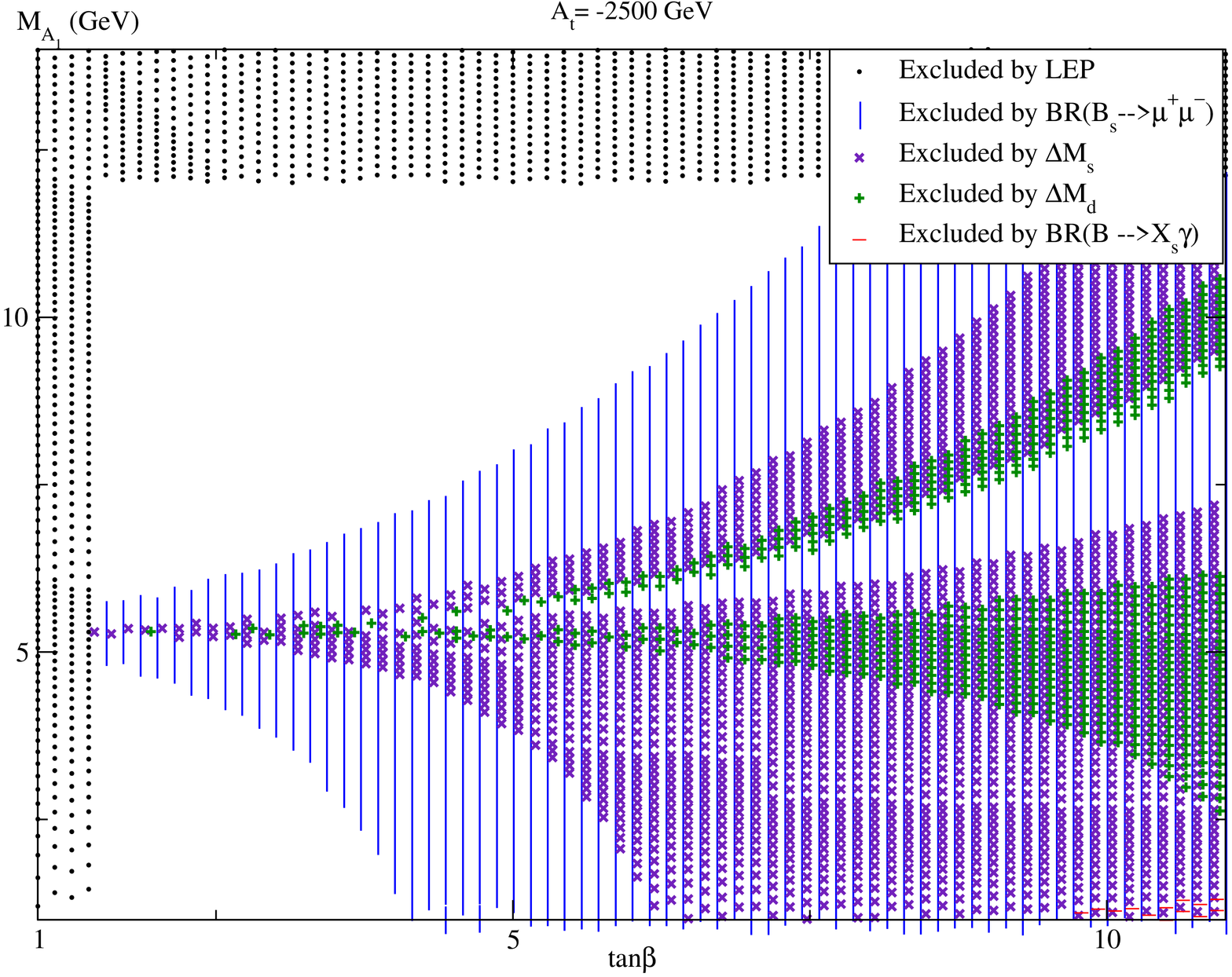}
\end{center}
\caption{Constraints in the $\tan\beta$-$M_{A_1}$ plane for
$A_t=-2500$~GeV.}
\end{figure}

Next we discuss a region specific to the NMSSM: In the NMSSM,
singlet-like pseudoscalars $A_1$ even below 10~GeV are able to
survive LEP constraints.
However, their loop induced flavour violating couplings to quarks and
leptons can be large enough to cause significant contributions to $B$
physics observables, most of all for CP odd scalar masses near the
resonance ($m_{A_1}\sim M_{B_q}$) and/or large $\tan\beta$. (Now, large
values of $\tan\beta$ do not only lead to larger couplings of the light
CP odd scalars, but also to an increase of their width which, in turn,
enhances their contribution via the s-channel Penguin diagram even for
masses a few GeV away from the resonance.)

\begin{figure}[t]
\begin{center}
\rotatebox{-90}{\epsfig{file=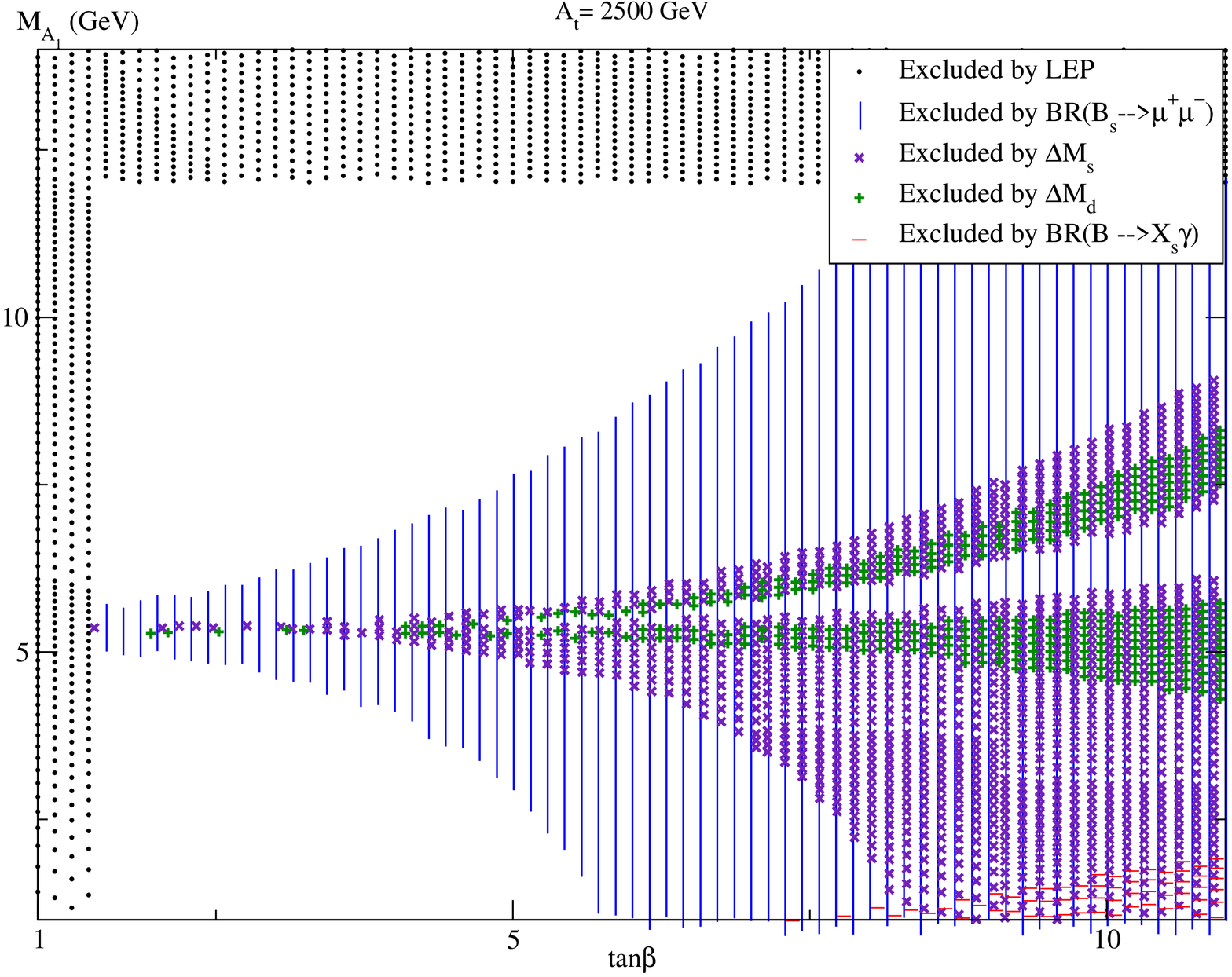,height=15cm}}
\end{center}
\caption{Constraints in the $\tan\beta$-$M_{A_1}$ plane for
$A_t=2500$~GeV.}
\end{figure}

In the following we present several examples of constraints on
$\tan\beta$ and $M_{A_1}$ that originate from $B$ physics processes.
For the soft Susy breaking squark and gaugino masses we take the same
values as above, and 300~GeV for $\mu_{eff}$. The NMSSM specific
parameters are chosen as $\lambda = 0.45$, $\kappa = 0.4$ and
$A_{\kappa} = -30$~GeV. However, $A_{\lambda}$ (or the MSSM-like
parameter $M_A^2 = \lambda S (A_\lambda + \kappa S)/(\cos\beta
\sin\beta)$) must be chosen within a relatively narrow
$\tan\beta$-dependent window such that LEP constraints on all CP even
and CP odd Higgs scalars remain satisfied.\break 
In Figs. 8--10 $M_A$ is chosen within this $\sim$1--2~GeV wide window.
($M_A$ varies from 300 to 400~GeV for $\tan\beta \sim 1.5$ to 10; LEP
constraints would allow to extend this window up to $\tan\beta=50$
with  more finetuning on $M_A$; however, B physics constraints exclude
this domain.)

\begin{figure}[t]
\begin{center}
\rotatebox{-90}{\epsfig{file=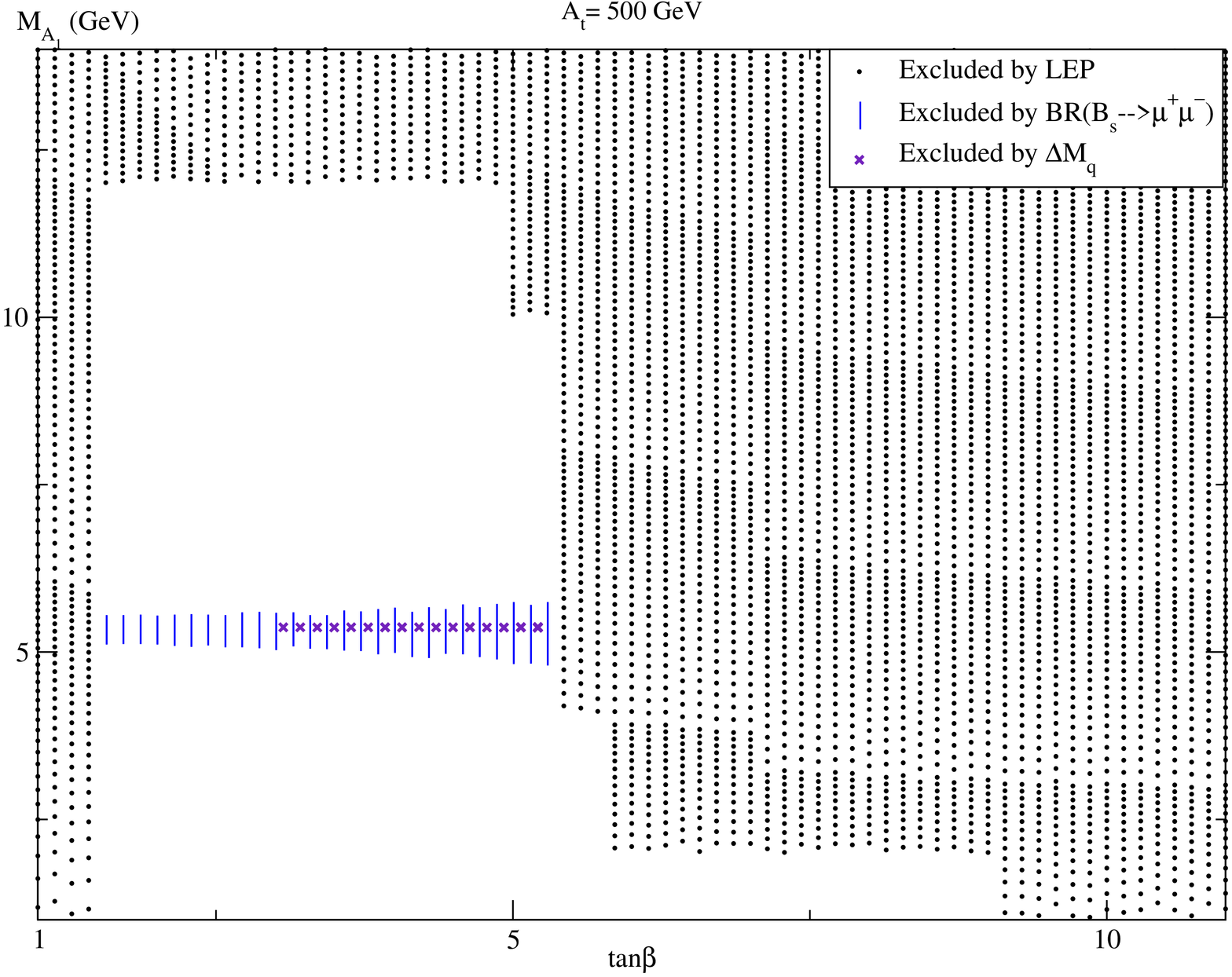,height=15cm}}
\end{center}
\caption{Constraints in the $\tan\beta$-$M_{A_1}$ plane for
$A_t=500$~GeV.}
\end{figure}

In Fig. 8 we consider the plane $M_{A_1}$ vs. $\tan\beta$, and assume
$A_t = -2.5$~TeV. Now, the constraints from  $BR(\bar{B}_s \to \mu^+
\mu^-)$ are the most relevant, and lead to strong upper limits on
$\tan\beta$ at fixed $M_{A_1}$. Among the remaining observables,
constraints from $\Delta M_s$ are also significant but generally 
redundant with the respect to $BR(\bar{B}_s \to \mu^+ \mu^-)$.

Whereas the situation for $A_t = 2.5$~TeV in Fig. 9 is similar to the
one with $A_t = -2.5$~TeV (the main difference comes from  $BR(\bar{B}
\to X_s \gamma)$, which excludes now a region with very light $A_1$ and
$\tan\beta \gsim 7$ already covered by $BR(\bar{B}_s \to \mu^+
\mu^-)$), the case $A_t = 500$~GeV considered in Fig. 10 is quite
different: In contrast to Figs. 8 and 9 the lightest scalar Higgs  mass
$m_h$ is below 90~GeV, but LEP constraints can still be satisfied due
to the decay $h \to A_1 A_1$. 

Note that, on the one hand, $A_1$ in Fig. 10 has a $\sim 90\%$ singlet
component, but also a $\sim 40\%$ doublet component. For $\tan\beta$
near 5 its coupling to down type quarks is even $\sim 2$~times larger
than the one of a SM scalar Higgs boson. As function of $M_{A_1}$ (and
$m_h$), LEP constraints on $h \to A_1 A_1 \to 4\ b$, $h \to A_1 A_1 \to
4\ \tau$ or $h \to A_1 A_1 \to 4$ jets have then to be applied, which
explains the jumps in the upper bound on $\tan\beta$. However, within
the region allowed by LEP, B physics constraints are particularly weak:
only a narrow stripe with $M_{A_1}$ near $M_{\bar{B}_s}$ is excluded
by  $BR(\bar{B}_s \to \mu^+ \mu^-)$ and $\Delta M_{s}$. Once again,
this is due to the fact that neutral Higgs effects are essentially
proportional to $|A_t|$ and small for small~$|A_t|$.

\section{Summary and Outlook} 

In this article, we have updated constraints from B physics observables
on the parameters of the MSSM and the NMSSM (assuming minimal flavour
violation), combining them with LEP constraints on the parameter space.
Available SM and BSM radiative corrections are included in the
calculations, which will be made public in the form of a Fortran code.

As expected, constraints from $BR(\bar{B} \to X_s \gamma)$ have become
weaker due to the recent increase of the world average, and the
decrease of the SM prediction (which is now below the experimental
average). Our numerical results (summarized in Figs. 1--4) show that
constraints still arise if, simultaneously, $M_{H^\pm}$ is small
($M_{H^\pm}\lsim 300$~GeV) and  $\tan\beta$ not too large ($\lsim 10$),
or if $\tan\beta \gsim 10$ and $|A_{t}|$ is large. We have verified
explicitely (for the first time), that NMSSM specific contributions to
$BR(\bar{B} \to X_s \gamma)$ are numerically negligible.

Among the other processes, $BR(\bar{B}_s\to \mu^+ \mu^-)$ is typically
the most sensitive and can exclude regions in parameter space for
$\tan\beta \gsim 15$ that would be allowed by $BR(\bar{B} \to X_s
\gamma)$, see Fig.~6. However, also $BR(\bar{B}^+ \to \tau^+ \nu_\tau)$
can lead to the most relevant constraints for very large $\tan\beta$,
cf. Fig.~7.

In the NMSSM specific case of a light CP odd Higgs scalar, constraints
from $BR(\bar{B}_s\to \mu^+ \mu^-)$ (inside the LEP allowed region) are
quite strong for large $|A_{t}|$ (cf. Figs. 8 and 9), but exclude only a
small region around $M_{A_1} \sim 5$~GeV for small $|A_{t}|$ (cf.
Fig.~10).

In the future, our calculations will allow to combine constraints from
B  physics observables with additional assumptions such as universal
soft terms at the GUT scale (the CMSSM and the CNMSSM) and/or
constraints from the dark matter relic density via
NMSSMTools~\cite{nmtools}.

\end{document}